**Chapter 5   AI in HCI Design and User Experience**

Wei Xu,  Ph.D.

(Dec. 2022)

Zhejiang University, China

## 5.1 Introduction

Recently, much progress has been made in artificial intelligence (AI) and machine-learning (ML); such progress has also enabled human-computer interaction (HCI) and user experience (UX) professionals to deliver solutions with better UX (Lu et al., 2022; Yang et al., 2018; Kuniavsky et al., 2017). The use of AI/ML capabilities for improving HCI/UX work and delivering better UX in solutions is becoming a trend (Abbas et al., 2022; Wu et al., 2019; Nikiforova et al., 2021) and creates many new opportunities for HCI/UX professionals (Holmquist, 2017; Yang et al., 2020). Some even speculate "AI/ML is the new UX" (Yang et al., 2018).

Researchers proposed that AI can perform as an assistant, collaborator, researcher, or facilitator (Bertão & Joo, 2021; Main & Grierson, 2020). AI technology will change the role of designers in the design process and generate an opportunity for creative collaboration between AI and designers (McCormack et al., 2020). Also, companies are moving fast to adopt AI for improving customer experience (CX). In 2018, the IBM Institute for Business Value (IBV) surveyed 1,194 executives from seven industries worldwide who are responsible for the AI initiatives of their companies (Schwartz et al., 2018). The results show that 74% said AI would fundamentally change how they approach CX; 41% had an AI strategy considering the changes ahead. To some extent, AI and UX designers have similar functions. They both gather data, analyze users' behavior and interactions, and can predict human behavior (Donahole, 2021). For example, Chatbots, Google Translate, and Alexa are good examples of AI technology that uses big data to deliver enhanced UX.

Table 5.1 summarizes the benefits of AI technology in enhancing HCI/UX design (e.g., Yang et al., 2018; Inkbot Design, 2021; Rogers, 2020; Schwartz et al., 2018; Baker, 2019; Donahole, 2021). Herein, AI refers to technologies (e.g., AI, ML, big data) used to develop AI-based intelligent systems, applications, or services.

**Table 5.1  Benefits of AI for HCI and UX design**

| Benefits of AI | Descriptions and Examples |
|---|---|
| Sensing users intelligently and supporting user research effectively | • Collect user personal knowledge (e.g., online shopping behavior, social connections)<br>• Recognize user's activity (e.g., physical status and interaction with systems)<br>• Infer the user's internal status (e.g., intention, emotion, attitude)<br>• Identify unique characteristics and behavioral patterns of collective users (e.g., digital personas)<br>• Sense context of user interaction (e.g., online historical shopping data) |
| Acting intelligently based on insights from sensing users | • React with appropriate actions (e.g., inform, engage, assist, promote products)<br>• React proactively or autonomously (e.g., empathy, influence, conflict management)<br>• Analyze and optimize customer journeys |
| Personalization design | • Provider personalized online content and functionality based on user preferences and interactions (e.g., data log, digital personas)<br>• Promote marketing experience to another level by using users' personal information<br>• Deliver advanced localization capabilities to handle language-related activities<br>• Focus on satisfying the precise needs of users |
| Analyzing a large amount of data more quickly and efficiently | • Reveal insights that help users rapidly make decisions by providing real-time responses to user inquiries<br>• Analyze large amounts of data to ascertain patterns and deliver meaningful research results (e.g., quickly generate questionnaires, and provide relevant responses for further inquiry)<br>• Process vast datasets to gather information<br>• Analyze massive sets of data to modify user experiences<br>• Simulate intellectual cycles to empower decision-making<br>• Automate mechanical errands at excellent paces<br>• Gather and draw inferences from vast volumes of data in a time |
| Powering HCI/UX activities for efficiency | • Automate repetitive design activities (e.g., resize images, make color corrections, crop images)<br>• Help generate a creative idea in the early design stages<br>• Leverage algorithms to create flow diagrams in UI design based on historical user patterns<br>• Develop wireframes based on the understanding of the context and the flow<br>• Help generate multiple variants of a UI design solution for A/B testing<br>• Automate design tasks, e.g., identify patterns in images and help designers stitch them together |
| | • Run dynamic A/B testing and analyze test results |
| | • Automate back-end processes (e.g., automate targeted marketing promotions) |
| | • Help designers quickly make design decisions (e.g., predictions based on historical datasets, giving users the fewest potential choices |
| Providing more natural and effective interaction | • Enable new types of user interface technologies (e.g., voice input, face recognition, gesture interaction, brain-computer interface) |
| | • Handle inaccurate input through reasoning (e.g., user intent detection, affective interaction) |
| | • Build thinner UI with AI (e.g., using historical data) to anticipate a user's action or better prioritize the queries of users, provide a possible solution or pertinent results |
| Better marketing | • Help build a better connection among brands across target audiences and boost their relationship. |
| | • Build customized eCommerce sites using their personal information, taking the marketing experience to another level. |
| Working as a user assistant | • Help predict user behavior for improving UX (e.g., Siri, Alexa) |
| | • Integrate into end user-facing solutions for users to interact with directly (e.g., chatbots, robots) |
| Working as a capability | • Work as an app service tailored towards a particular action or use case (e.g., search) |
| | • Provide ML-based speech-to-text services |
| | • Provide analytic capabilities (e.g., risk assessment, sentiment analysis, retroactive analysis) |
| | • Provide text-related capabilities (e.g., natural language processing, text recognition, speech-to-text conversion) |
| | • Provide visual capabilities (e.g., computer vision, augmented reality) |

Thus, it is apparent that AI is transforming how HCI and UX professionals work towards delivering optimal UX in their solutions. The transformation impacts the HCI/UX activities such as user research, user interface (UI) technologies and design, and user evaluation.

In this chapter, we review and discuss the transformation of AI technology in HCI/UX work and assess how AI technology will change how we do the work. We first discuss how AI can be used to enhance the result of user research and design evaluation. We then discuss how AI technology can be used to enhance HCI/UX design. Finally, we discuss how AI-enabled capabilities can improve UX when users interact with computing systems, applications, and services.

## 5.2 AI in HCI/UX research and evaluation

### 5.2.1 Overview

The goal of HCI/UX research and evaluation is to systematically gather and analyze user data through HCI/UX activities (e.g., user research, usability testing) to understand a problem space (e.g., user pain points, usability issues) and guide the entire design process (Xu, 2005). Conventional approaches to user study rely on methods such as surveys or user interviews; for UX evaluations, HCI/UX professionals manually conduct usability testing of a proposed design to identify issues and then analyze data to generate recommendations for design improvement (Xu, 2017). However, these methods are time-consuming.

A few years ago, researchers found that there is only little academic work at the intersection of UX and AI (Chromik et al., 2020; Yang et al., 2018); they found even less research explicitly addressing AI/ML for user research. However, the number of relevant publications has been increasing since 2015 and continues to do so as AI is gaining popularity in many contexts.

The first AI-based approach uses big data-driven user research methods that gradually replace traditional methods. With the development of the 5G, the Internet of Things, smart devices, etc., user data generation inevitably grows explosively. Big data technology adoption in user research shows an upward trend. However, the way of collecting user data is through some AI technology (e.g., sensing, face recognition) and user interactions (e.g., user interaction log, online click streams, and social media); many of the big data-driven approaches are based on traditional statistical techniques have not fully leveraged AI methods in their analysis for the collected user data (Dan et al., 2020).

As an alternative approach, ML-based approaches are primarily used to support the analysis of already collected user data (Chromik et al., 2020). For example, it analyzes textual user data. ML and natural language processing (NLP) methods have been used to semi-automate the coding of interview transcripts (Marathe et al., 2018) and to extract UX-related problems from online review narratives through classification (Mendes et al., 2017). Data-driven learning approaches have also been used to construct behavioral personas from user interaction log data.

The third approach combines the two methods discussed above, applying AI/ML technology to both data collection and analysis stages. For example, Gartner (2018) provides recommendations for customer journey analytics (CJA), leveraging existing analytics tools (when available) and incorporating ML as a service (i.e., MLaaS) products to enhance analytical capabilities toward CJA. MLaaS incorporates ML-driven recommendations to power customer journey orchestration solutions and enhances digital experiences, optimizing customer journeys through Interaction Point Analysis across crucial interaction points.

Table 5.2 summarizes the main benefits of AI/ML across HCI/UX research and evaluation activities (e.g., Chromik et al., 2020; Delrieu et al., 2018; Baker, 2019).

**Table 5.2  The benefits of AI/ML across HCI/UX research and evaluation activities**

| HCI / UX Activities | Benefits |
|---|---|
| | |
| Data collection | • Engaging surveys: simplify survey studies by leveraging the idea of adaptive user interfaces (i.e., questionnaires might automatically be tailored in real-time to the individual survey participant based on their previous answers)<br>• Remote tracking of user behavior over time<br>• Applying conversational and voice user interfaces for more empathetic survey studies |
| Evaluation of design | • Data-Driven Design: Supporting design decisions by evaluating and recommending UI options based on historical data of user behavior or user preferences was considered another field of interest. |
| Analysis | • Analysis of ordinal data (e.g., questionnaires)<br>• Analysis of text (e.g., transcripts of user research/evaluation)<br>• Analysis of voice/audio-based data (e.g., recordings, camera feed)<br>• Analysis of log data (e.g., click behavior)<br>• Analysis of time series data (e.g., mouse and eye movement)<br>• Analysis of emotion and sentiment<br>• Automated transcription: ML-based speech-to-text services for the post-processing of contextual inquiries or interviews<br>• Excel at quickly analyzing vast amounts of existing data to identify subtle patterns in dispersed data silos and to inform UX insights<br>• Detection of patterns within structured or unstructured data<br>• User modeling<br>• Augmented/predictive analytics: Insights automatically generated from ML bring up new opportunities in conversion funnels |
| Generation of UX artifacts | • User personas<br>• Customer journeys: Leveraging ML functionality allows organizations to contact or re-engage existing or potential customers at the optimal time in the customer journey and through the optimal communication channel<br>• Audience/customer segmentation: By quickly analyzing large datasets and identifying patterns, these tools help analysts discover and validate new customer cohorts or segments |

### 5.2.2 Digital personas

One practical approach to using the data from **user research** is to develop **personas** (Alan Cooper's theory). **Persona refers to a** group of users with similar behaviors and goals within the group and with differences in behavior between the groups. Personas allow HCI/UX professionals to identify and understand the differences in how a product is used by different groups of users based on their usages and behaviors so that the product can provide a tailored experience (e.g., functions, content) across personas. Generating personas can become challenging and time-consuming in conventional user research, especially when HCI/UX researchers need to interview many users and plan to build many personas. Also, a manual process of developing personas has been criticized for creating personas that are not based on rigorous empirical data. The process often uses small samples, one-time data collection, and non-algorithmic methods (Salminen et al., 2021).

The AI era generates a vast amount of user data, reflecting the user's behavior, preferences, and demands, and has high research value in building personas. Data-driven personas, called "digital personas," have gained popularity in HCI due to digital trends such as personified big data, online analytics, and the evolution of data science algorithms. Specifically, three trends have significantly transformed the way how we build personas (Tan et al., 2020): (1) availability of user data from online analytics and social media platforms; (2) democratization of data science tools and algorithms that enable automated persona generation; and (3) Web technologies that remove the limitations of static personas via interactive user interfaces. These three trends allow us to use algorithmic methods to create accurate, representative, and refreshable personas from numerical data (Salminen et al., 2020, 2021).

The benefits of digital personas are apparent. AI/ML-based algorithms allow HCI/UX professionals to develop personas much faster than conventional manual processes. An algorithm-based analysis of collected user data also can help identify distinct personas and visualize how HCI/UX professionals reliably identified their attributes. For example, Salminen et al. (2022) introduced Persona Analytics (PA), a system that tracks how users interact with data-driven personas. PA captures users' mouse and gazes behavior to measure users' interaction with algorithmically generated personas. The researchers also conducted a study with 144 participants, demonstrating how PA could be deployed for remote user studies.

Salminen et al. (2021) summarize the distinct relative benefits of digital personas:

- Enhanced objectivity. Digital personas tend to be replicable, and they use large sample sizes to increase the user representativeness of the personas. The conventional manual process for developing personas is associated with a high degree of subjectivity, which hinders the validity of the created personas. The statistical robustness of digital personas boosts both the validity and credibility of the developed personas.
- Decreased cost. The manual process of creating personas is time-consuming and costly. Digital personas mitigate this cost by relying on automation in persona creation, including data collection and analysis, thus offering ways to "democratize" persona development for organizations of all kinds.
- Updatability. Shifts in user demographics and behaviors are typical in many fast-moving industries, such as Web-based businesses. Updating personas requires a high cost for the manual process, resulting in outdated personas. Digital personas can capture the change in user behavior over time based on their automated processes for systematic data collection and easy re-analysis using standard algorithms.
- Scalability. Manual data analysis is costly and requires specific expertise, making the personas built using manual processes less compatible with large datasets. Large datasets are common with social media and Web analytics and are not a concern for digital personas, as data science and ML algorithms have been developed to process large amounts of data

More and more researchers are analyzing massive data and data characteristics to generate digital personas. For example, Yu (2014) combined used tags to describe user characteristics. They extracted relevant information tags through clustering in the big data environment to present the complete picture of users through digital personas. Joni Salminen of Qatar Research Institute integrated data from Facebook, Twitter, YouTube, and generated real-time personas based on user profiles and interactive behaviors, which provides users with competitive marketing methods and strategies across different platforms (Salminen et al., 2017). From the perspective of cultural differences, An et al. (2016) analyzed millions of content from the Middle East on YouTube to generate personas for deeply analyzing cultural diversity's impact on users' social media use.

People have also attempted to build accurate personality traits and types for their target users as the foundation for building personas (Salminen et al., 2020; Sun et al., 2018; Carducci et al., 2018). The automatic assessment of personality dimensions relies on information gathered from social media platforms, such as a list of friends and interests in music and movie endorsements. The work turned the collected data into signals as inputs. Supervised ML approaches have been efficient and accurate in computing personality traits and types. Specifically, Carducci et al. (2018) proposed a supervised ML approach to define personality traits by relying on what an individual tweeted about publicly. The approach segments tweets in tokens and then learns word vector representations as embeddings that are then used to feed a supervised learner classifier. This study demonstrates the approach's effectiveness by measuring the mean squared error of the learned model to compare it with an international benchmark of Facebook status updates. Also, the study tested the transfer learning predictive power of the proposed model with an in-house built benchmark created by twenty-four panelists who performed a state-of-the-art psychological survey. The comparison shows that the proposed model received an excellent conversion while analyzing the Twitter posts towards the personality traits extracted from the study.

Salminen et al. (2020) built social media-based personas with personality traits based on a deep ML approach. They developed a deep learning classifier (a NN classifier) that predicts personality traits. The study used three publicly available datasets and applied an automatic persona generation methodology to generate 15 personas from the social media data of an online news platform. After developing the personas, the study aggregated each persona's YouTube comments and predicted the personality traits of each persona from the comments on that persona. The results indicate an average performance increase of 4.84% in scores as compared with a baseline.

However, there are challenges for digital personas (Salminen et al., 2021). The main challenges include (a) data quality; (b) data availability; (c) method-specific weaknesses, such as the accuracy of the algorithm, people behaving differently across different segmentations; (d) human and machine biases, such as the persistent need for judgment calls ("manual labor") that creates a potential source of bias and obstacles for completely automated digital personas; (e) validation of digital personas methods; (f) lack of standardization; and (g) lack of consideration for inclusivity.

### 5.2.3 Qualitative analysis

In user research and UX evaluation, HCI/UX professionals must conduct a qualitative analysis of the collected data, such as interview transcripts in text or recorded video files (natural language), usability test video recordings, and social media data. Qualitative analysis involves identifying themes, grouping data points based on these themes, and establishing relations between these groups. This type of qualitative analysis is time-coming and subjective to some extent. Many HCI/UX professionals have considerable training in analyzing human behavior while users interact with computing systems. Still, AI applications for this kind of behavioral data have yet to leverage their expertise fully. AI technology provides one approach to help the qualitative analysis for HCI/UX professionals. Therefore, developing and leveraging AI-assisted capabilities for qualitative research is critical.

The application of AI in quantitative analysis has been increasingly popular over the last several years. For example, researchers proposed using natural language processing (NLP) and ML to generate initial codes followed by humans correcting the codes; other work utilized NLP to derive potential codes and models (Chen et al., 2018). Although progress has been made in developing AI-assisted capabilities for qualitative analysis, there are still challenges, and low accuracy has been considered the primary limitation of such automated approaches. Chen et al. (2018) argued that we use AI/ML to support qualitative coding for identifying ambiguity. HCI/UX activities generate petabytes of free-form text data, recording daily user experiences; NLI (natural language inference) allows analysis of this large-scale data, but NLI used to be applied for analysis with organized datasets, and less is known about how to design NLI for querying and analyzing text data (Mishra & Rzeszotarski, 2020).

Specifically, Liu et al. (2020) deployed a semantic data analysis processing approach. The approach introduced a specific implementation method using AI/ML semantic analysis technology to analyze language materials in user research. The effectiveness of applying the AI semantic analysis techniques in user research was tested and verified. The result shows that this application of AI technology has demonstrated the potential that AI technology can replace some of the human manual analysis tasks for efficiency improvement.

User intent analysis is also essential in user research and UX evaluation. A method called Natural Language Interaction (NLI) is to do user intent analysis (Setlur, 2020). For example, NLI was applied to a labeled dataset that captures user intent distribution, co-occurrence, and flow patterns. Specifically, Setlur (2020) employed deep ML techniques that approximate the heuristics and conversational cues for continuous learning in a chatbot interface. These data-driven approaches help broaden the scope for visual analysis workflows across various chatbot experiences.

Analyzing usability test videos is also challenging for HCI/UX professionals. Fan et al. (2022) have explored how AI can help facilitate effective collaboration between UX evaluators and AI. Based on the previous work in human and AI agent collaboration, they studied two primary factors: explanations and synchronization. Explanations allow AI to inform UX professionals how it identifies UX issues from a usability test session; synchronization refers to the two possible ways UX professionals and AI collaborate: synchronously and asynchronously. By adopting a hybrid wizard-of-oz approach to simulating an AI solution with good performance, they conducted a mixed-method study with 24 UX evaluators who were asked to identify UX issues from usability test videos using the AI-assisted capability. The results show that AI with explanations, whether presented synchronously or asynchronously, provides better support for UX evaluators' analysis; when without explanations, synchronous AI better improved UX evaluators' performance compared to asynchronous AI. This study also implies that an AI-assisted UX evaluation can facilitate more effective human-AI collaboration.

Analyzing the structure of texts is an alternative way for qualitative analysis. Recently, personality detection based on texts from online social networks has attracted more and more attention. Sun et al. (2018) analyzed texts' structure as an additional dimension for practical qualitative analysis. Previous models were based on letters, words, or phrases, which is insufficient to get good results. Sun et al. (2018) present a preliminary research result that shows the structure of texts can also be an essential feature in studying personality detection from texts. More specifically, the study deployed a model called 2CLSTM. 2CLSTM is a bidirectional LSTM (Long Short-Term Memory network) concatenated with CNN (Convolutional Neural Network), which can detect a user's personality based on text structures. They conducted evaluations across two datasets containing long and short texts. The results have achieved better results, demonstrating the proposed model can efficiently learn useful text structure features for qualitative analysis.

Chen et al. (2018) highlight two challenges for ML applications in qualitative coding. On the one hand, a lack of understanding between disciplines may negatively impact trust, limiting the application of ML in qualitative analysis because people using qualitative methods are generally not trained in ML techniques. In some cases, ML experts' limited understanding of social science values and methods can hamper effective collaborations. On the other hand, there are fundamental differences between qualitative and quantitative methods. In quantitative analysis, data points that appear very few times may be considered noise, but from a qualitative analysis perspective, the quantity of instances does not always reflect significance.

As for future work supporting qualitative analysis in HCI/UX activities, we anticipate that more human-centered and interpretable AI methods can potentially transform social science research (Chen et al., 2018). Specifically, current AI models are not always interpretable, and the AI community needs to increase the transparency and interpretability of AI technologies for qualitative analysis. Also, we need to explore ways to make the usage of AI-based capabilities a meaningful task in HCI/UX practices, bridging the gap between AI and HCI/UX communities.

### 5.2.4 UX evaluation

HCI/UX professionals conduct UX evaluations to achieve design improvements based on user feedback and insights. However, traditional UX evaluation methods like questionnaires and usability testing are often resource-intensive and not scalable. With the continuous evolution of many intelligent products (e.g., AI assistants, autonomous driving, smart homes, intelligent robots), these UX evaluation methods will be difficult to meet new scenarios (Lan et al., 2020). Emerging technologies such as AI and big data are currently influencing how HCI/UX professionals conduct UX evaluations (Lan et al., 2020; Tan et al., 2020).

The methods of finding products with poor UX through big data-based analysis are gradually being adopted by collecting the user's stay time, login frequency, conversion rate, and other indicators (Tan et al., 2020). For example, Yu (2018) designed a big data intelligent algorithm framework for UX evaluation of mobile media clients through indicators such as the number of fans, page views, activity, stickiness, and emotional inclination, and finally implemented a cognitive algorithm framework. Li (2019) conducted an in-depth analysis of the user stickiness of NetEase Cloud Music through big data analysis algorithms designed for popularity. Recent work has also attempted to address other aspects of human behaviors on user interfaces, e.g., predicting human perception of UI interactivity based on user behavior data such as mouse/keyboard logs, eye tracking, and usage log (Swearngin & Li, 2019). Souza et al. (2022) establish a framework that employs eye and mouse tracking methods, keyboard input, self-assessment questionnaires, and AI-based algorithms to evaluate UX and categorize users in terms of performance profiles.

Furthermore, recent academic work explores the challenges of evaluating UX using multiple data sources and proposes ML-based approaches (Asim et al., 2020). Connecting questionnaire results with log and time series data about user behavior may be used as labeled data for input data to supervised ML. Such approaches may allow for continuous monitor changes in users' UX and inform HCI/UX professionals of the opportunity for improvement. Chromik et al. (2020) propose that some equipment, such as electroencephalography (EEG) sensors, could be used during real-time usability tests in lab contexts to record typical flows of interaction and users' emotional responses. These behavioral and emotional responses could be used as labels for an ML model (Chromik et al., 2020).

For usability testing, ML was used for selecting participants for usability tests (Gilbert et al., 2007) and A/B tests (Kharitonov et al., 2017). In addition, automatic real-time evaluation of mobile-based experience via emotional logging systems using video-captured facial expressions in lab contexts (Filho et al., 2015), using acoustic data (Soleimani et al., 2017) and skin conductance signals (Liapis et al., 2015).

Studies show that AI technology may offer a more resource-effective approach (Chromik et al., 2020). For example, Yang et al. (2020) proposed a methodology for measuring UX using AI-aided design (AIAD) technology in mobile application design. AIAD focuses on the rational use of AI technology to measure and improve UX. The researchers propose to obtain user behavior data from logs of mobile applications. They designed and used projected pages of the application to train neural networks for specific tasks in terms of the click information of all users when performing the tasks. The goal was to make the deep neural network model simulate the user's experience in operating a mobile application as much as possible. Thus, user behavior features could be aggregated and mapped in the connection and hidden layers. Finally, the optimized design was executed on the application to verify the efficiency of the proposed methodology.

We further provide two more examples illustrating how AI technologies can help facilitate UX evaluation with three different approaches: visual search performance modeling, user emotion detection, and user interaction behavior modeling.

The first example involves visual search performance modeling (Yuan et al., 2020). Modeling visual search performance not only offers an opportunity to predict the usability of an application before actually testing it on real users but also helps HCI/UX professionals better understand user behavior. The authors first analyzed a large-scale dataset of visual search tasks on actual web pages. They then presented a deep neural network that learns to predict the scannability of webpage content, i.e., how easy it is for a user to find a specific target. The model leveraged heuristic-based features such as target size and unstructured features such as raw image pixels. The model then analyzed the user behaviors to offer insights into how the salience map learned by the model aligns with human intuition and how the learned semantic representation of each target type relates to its visual search performance. This approach allows HCI/UX professionals to model complex interactions involved in visual search tasks, which traditional analytical methods cannot quickly achieve.

The second example is user emotion detection. Emotion is one aspect of UX that exploits an ML-based automatic UX evaluation for understanding users' emotions by analyzing the log data of the users' interactions with websites (Desolda et al., 2021). The evaluation results show the performance of each ML algorithm according to the seven emotions. It is evident that emotions like sadness, anger, fear, disgust, and surprise were predicted with higher accuracy; joy was instead predicted with medium accuracy, while contempt had lower accuracy in all cases.

Lastly, Bakaev et al. (2022) deployed neural networks-based approaches for predicting the visual perception of UI. As testing and validation of graphical user interfaces (GUIs) increasingly rely on computer vision, convolutional neural networks (CNN) models that predict UX start to achieve decent accuracy. However, CNN models require vast amounts of human-labeled training data, which are costly or unavailable for HCI/UX activities. This study compares the prediction quality of CNN and artificial neural networks (ANN) models to predict visual perception in terms of aesthetics, complexity, and orderliness scales for about 2700 web UIs assessed by 137 users. The results suggest that the ANN architecture produces a smaller mean squared error (MSE) for the training dataset size (N) available in our study but that CNN should become superior with N > 2912.

While using AI technology in UX evaluation is promising, we still face challenges (Li et al., 2020). For instance, deep ML methods are often data-hungry, while interaction data is relatively scarce compared to classic ML problems such as computer vision or natural language processing. Deep ML models are not easy to analyze. While better modeling accuracy is of great benefit, the interpretability of a model is crucial for HCI/UX professionals to gain more insights about UX. Further collaborative work is needed between the AI and HCI/UX communities.

## 5.3 AI in HCI/UX design

### 5.3.1 AI for UI design

Designing an excellent GUI requires much innovation and creativity, but the process is time-consuming and error-prone (Lu et al., 2022). Recently, AI-driven design (e.g., algorithmically powered tools) has become popular. AI-driven design vows to move UX design to another degree of digital experience in support of wireframing automation, visual design analysis, and UI pattern-driven design (Baker, 2019). Many researchers have worked on building design support tools to improve the efficiency of UI work. Also, many commercial design prototyping tools are developed. These tools have greatly helped designers create UI prototypes supporting UX work.

Many initial AI-based capabilities were released to support UI design. For example, Adobe's Creative Cloud software can realize the intelligent analysis function of multimedia files such as images and videos. They can provide intelligent material recommendations according to the designer's design needs. Autodesk's Dream Catcher can quickly generate thousands of design proposals for designers to choose. Google released a new AI-based Alphard (Alpha Graphic design) with the output of high-quality graphic design solutions. Microsoft and Airbnb experimented with converting paper sketches directly into GUI code, bypassing much of the digital wireframing phase (Wilkins, 2018). Some of the tools are even beyond GUI. They utilize pre-trained AI algorithms that have the potential to support new forms of interaction by processing eye, face, body, and hand movements captured through webcams, speech commands captured through the browser's audio channel, and text through web elements (Li et al., 2020).

There are many (or potential) benefits to leveraging AI technology in UX design. Figure 5.1 illustrates how an AI-based approach could help remove some redundant work from human designers and developers from a design process perspective. The research presents some inspiring illustrations of ML-based UX design tools. For example, Gajjar et al. (2021) proposed Akin, a UI wireframe generator that uses a fine-tuned SAGAN model to generate UI wireframes for smartphone UI design. The researchers annotated and classified 500 UI screens from RICO into five commonly used mobile design patterns. The SAGAN model was trained with the dataset to generate UI wireframes for a given UI design pattern. An evaluation of Akin conducted with 15 UX designers shows that the designers rated the quality of wireframes developed by Akin as approximately equal to designer-made wireframes. Also, the designers could not distinguish UI wireframes generated by Akin from designer-made wireframes 50% of the time.

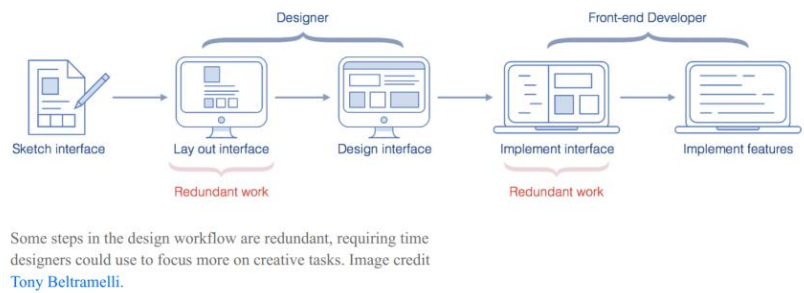

Figure 5.2  Illustration of an AI-based approach for UI work (Babich, 2020)

AI-based design capabilities also support design creativity, such as Simon's optimization-based design (Yang, 2017), which can all be linked to today's interest in employing ML to assist human creativity. TensorFlow.js is an open-source AI platform for developing, training, and using models in a browser or anywhere Javascript can run (Li et al., 2020). At Google, TensorFlow.js has been leveraged as a platform for AI + HCI collaborative research. TensorFlow.js is a browser-based ML framework to enable new forms of HCI/UX design innovation. TensorFlow.js provides a rich set of features accessible to researchers with different levels of ML experience. The library allows design experts to build models from scratch but also makes it easy to integrate pre-trained models.

The following list further summarizes some benefits (or potential benefits) of using AI-based capabilities supporting UI design (Baker, 2019; Lu et al., 2022; Vetrov, 2022; Chen et al., 2018; Abbas et al., 2022):

- Quickly make various design varieties per the user's response.
- Support design creativity
- Make wireframing and prototyping work more efficiently and less monotonously
- Transform UI sketches directly into a prototype
- Potential to immediately change a whiteboard sketch over into a functional prototype

- Quickly design alternative exploration
- Support design customization to support personalized UX
- Do design guideline violation check
- Empower design decision-making
- Help prepare UI assets and content
- Translate digital UI mockups into UI specifications

A representative approach of AI-based UI design is called *Generative UI Design*. Generative UI design is based on AI generative technology. Generally speaking, with the development of AI generative technology, people can realize collaborative creation with AI in music, painting, writing, design, dance, etc. (Li et al., 2020). AI generative technology can quickly generate new samples that meet the specifications based on specific data sets so that novices can quickly start creation or reduce the repetitive work of designers. For example, analyzing big data on clothing design through modeling and visualization techniques to expand the ideas of clothing designers and gain insights (Glauser et al., 2019).

With the millions of websites and mobile apps available, many UX problems an HCI/UX designer encounters may have already been considered and solved by someone else. Specifically, in the UI design area, generative models (e.g., Variational Autoencoders) were trained on a large set of UI design examples that can suggest design alternatives for HCI/UX designers (Li et al., 2020). Systems based on these AI methods often leverage human support or" Wizard of Oz" techniques to collect the data from large design samples and eventually generate design solutions informed by the collected data (Vaccaro et al., 2018).

Researchers have promoted a hybrid intelligence approach for effective generative UI design. Specifically, rather than using humans solely for data collection to train an AI system, the hybrid intelligence approach incorporates human users, often crowd workers, as an essential and permanent component in an interactive system for complex design tasks (Lasecki, 2019; Li et al., 2020). Such a hybrid intelligence approach provides rich opportunities to combine human and machine intelligence to collaborate on a task and improve each other dynamically and interactively. A system powered by the hybrid intelligence approach needs to synthesize responses from multiple designers to achieve acceptable performance or availability for the system.

Another approach for effective generative UI design is to foster a creative, generative ML approach (Kayacik et al., 2019). Research shows that such an approach is more robust when multiple designers with different points of view actively contribute to them. Currently, many UXers do not have the ML education needed in the industry. This lack of education is hampering ML research teams' capacity to have a broad impact on their projects. To address the issue, the Google People and AI Research (PAIR) group developed a novel program method in which UXers are embedded into an ML research group for three months to provide a human-centered perspective on the creation of ML models. The first full-time cohort of UXers was embedded in a team of ML research scientists focused on deep generative models to assist in music composition (Kayacik et al., 2019). At the end of three months, the UXers had new ML knowledge, and ML research scientists had a greater understanding of user-centered practices. The PAIR program results show that UX research and design involvement in creating ML models help ML research scientists more effectively identify human needs that ML models will fulfill.

However, the generative UI design method is limited, so the design realization is still limited to the traditional UI visualization level (Xu & Ge, 2018). It attaches great importance to the novelty of the appearance but lacks attention to UX design. It is not mature enough to deliver optimal UX to HCI/UX designers, especially with the business processes and structure built. Researchers also found that the innovative algorithm's information overload and uncertain output in human-AI collaboration are the key challenges (He et al., 2019). Also, AI algorithms will produce inaccurate judgments but lack explanations, reducing the user's perception of them (Glauser et al., 2018). Morris et al. (2022) proposed two design spaces for consideration when developing future generative AI models: how HCI can impact generative models (i.e., interfaces for models) and how generative models can impact HCI (i.e., models as an HCI prototyping material).

### 5.3.2 AI as a new design material

As AI technology advances, HCI/UX professionals regularly integrate AI capabilities into new apps, devices, and systems (Dove et al., 2017). AI becomes available as a resource to use by non-experts like HCI/UX professionals. First and foremost, intelligence is becoming a new design material (Holmquist, 2017). The options of a designer are, to a large extent, defined by the materials they have to work with. For instance, a product designer would need to be aware of the physical characteristics of materials such as plastic, wood, and metal, as well as how these fit together mechanically, to design an aesthetically and functionally pleasing experience. As AI becomes a more vital part of everyday products, HCI/UX designers will have to figure out how to work with intelligence as a new material.

With AI as a new design material, the primary role for HCI/UX designers is currently transitioning to augment end users with extended capabilities (e.g., new ideas, emotion design), besides routine UI design work. HCI/UX designers are becoming creators of the interface between humans and technology by leveraging algorithms. AI as a design material means that HCI/UX designers should view AI as a capability as an application service tailored to a specific functionality to support a particular experience (e.g., a "search" function). For instance, for an application with conversational UI. The traditional approach is that a customer asks a question, and a human agent responds for help. If we add AI to that conversation function by training models of the language, having those models process that language and algorithmically build the best response and return that response with an AI-based virtual support agent. Thus, this new material of invisible, personalized, conversational design is algorithms. HCI/UX designers can take an active role in bridging algorithms and the UI to bring significant experience to end users with technology.

AI as a new design material is essentially an algorithm as a new material. However, algorithms have many limitations, impacting the outcome of using these "design materials." Pavliscak (2016) lists several limitations of algorithms:

- Algorithms are not neutral or objective. Algorithms have a point of view with potentially biased outputs. Humans create algorithms, so their point of view gets embedded in the system
- Algorithms don't understand you as a complex individual. Algorithms generalize, simplify, and filter out

things they consider to be irrelevant.

- In many cases, algorithms use other people's data to fill in missing bits and pieces. The result is that algorithms don't reflect complicated humans.
- Algorithms are opaque. It's not always clear how or why they work the way they do. People who write them don't fully understand how they work.

There has been continuing study on how to approach UX design practice while working with AI as a design material (Dove et al., 2017; Yang et al., 2018; Amershi et al., 2019). Researchers argue that while data tell us about people and organizations, algorithms create guidelines, and ML shapes the experience (Pavliscak, 2016). Algorithms are considered a set of guidelines on how to perform a task. When you send a text message or do an Internet search, HCI/UX triggers a nested set of interdependent algorithms. To best make use of algorithms for UX, Holmquist (2017) proposed the following design guidelines when algorithms are used for design:

- Reveal the effects of algorithms: users don't understand how algorithms work, and experience designers need to make algorithms' results more apparent
- Participatory design for algorithms: let users participate in their data creation for a personalized experience and choose a level of trust using different personal preferences
- Designing for transparency: let users understand how AI affects their interaction with applications
- Designing for opacity: it is no longer possible to explain exactly why or how an AI does what it does
- Designing for unpredictability: no matter how well-trained a neural network is, it is still drawing its conclusions from given data.
- Designing for learning: the learning must be built into the interaction and completely unobtrusive, so it does not feel like the user doubles as the AI's training wheels.
- Designing for evolution: AI systems will continue to evolve. It will be necessary to communicate this to users so that they know what to expect and can benefit while avoiding unpleasant surprises
- Designing for shared control: how AI systems can be designed to allow the sharing of power with users

While AI technology has brought in values for HCI/UX design, we still face challenges to fully leverage this new type of design materials. Dove et al. (2017) conducted a survey that shows some significant challenges: (1) There are challenges with using AI capabilities from a human-centered perspective. Current HCI/UX design education cannot prepare future design graduates to incorporate AI into their work. (2) While ML pushes the boundaries of design, the balance of collaboration with engineers and developers is currently such that design-led innovation is still rare. (3) UX/UI prototyping with AI/ML is difficult. HCI/UX designers used to create prototypes in the form of sketches, plans, and physical models made of paper, cardboard, or foam (Hallgrimsson, 2012).

### 5.3.3  AI as a design collaborator

Design ideation is a source of innovation in the early stages of a development process. Beyond the AI capability to support UI design as discussed above, we also need to rethink the current role of HCI/UX designers. UI design should be a creative process involving multiple iterations of different prototyping fidelities to create a UI design. With the help of AI, we need to consider how to enable AI to perform repetitive tasks for the designer while allowing the designer to take command of the creative process. This approach would greatly benefit designers in co-creating design solutions with AI (Liao et al., 2020). Such a collaborative creation with AI may further promote optimal experience in solutions (Oh et al., 2018).

Researchers have been promoting this approach. For instance, Liao et al. (2020) proposed a framework of AI-augmented design support for the early stages where AI's role in creativity is related to creating representation, triggering empathy, and promoting engagement. Similarly, McCormack et al. (2020) characterized AI as a creative agent system that provokes, challenges, and enhances human creativity. Verganti et al. (2020) further claimed that AI reinforces design principles such as human-centered design, leading to potentially more creative solutions. AI will enable the designer's work, boost their creativity, and help experts create the best quality design products in a minimum time (Inkbot Design, 2021).

Main & Grierson (2020) proposed that AI can perform as an assistant, collaborator, researcher, or facilitator but might also play the role of future co-creator. Furthermore, McCormack et al. (2020) consider AI a system that allows creative collaboration with designers. Li et al. (2020) argued that rather than using humans solely for data collection to train an AI system, hybrid intelligence incorporates human users as an essential and permanent component in an interactive system for complex design tasks.

De Peuter et al. (2021) challenged the current approach and argued that AI for supporting designers needs to be rethought. It should aim to cooperate, not automate, by supporting and leveraging the creativity and problem-solving of designers. How to infer designers' goals and help develop a creative design needs to be figured out. They believe there is an urgent need to develop AI methods to cooperate with designers, working as assistants communicating with a designer about the design goal while supporting them in working toward that goal. In such a collaborative process, HCI/UX designers should remain the primary actor in the design process. As active participants, the designers explore and try things out to refine their goals. Further, the collaborative process can leverage the designer's creative abilities and expertise to build innovative designs.

Specifically, De Peuter et al. (2021) proposed a general-purpose approach for cooperative assistants in design problems. Collaborative design assistance has been offered for specific design problems, but the proposed method is to support a wide range of interactions. It uses a generative user model to infer a designer's goal from their behavior and plan how to assist the designer best. De Peuter et al. (2021) demonstrated the approach in a trip planning example (see Figure 5.2). As illustrated in Figure 5.2, AI should appreciate the explorative character of designers' thinking. Within this design process, designers generate solutions not only to solve a problem but also to learn about it, including its objectives and constraints. They can mentally plan over a design space based on a utility function (shown as contours in Figure 5.2). The utility function evolves as the design progresses. The AI should collaborate in this creative process, for example, by proposing high-quality solutions and complementing the designer's problem-solving. To do so, it needs to know the designer's utility function. The study suggested creating AI assistants (shown in blue) that can infer this utility from observations and then use it to assist a designer.

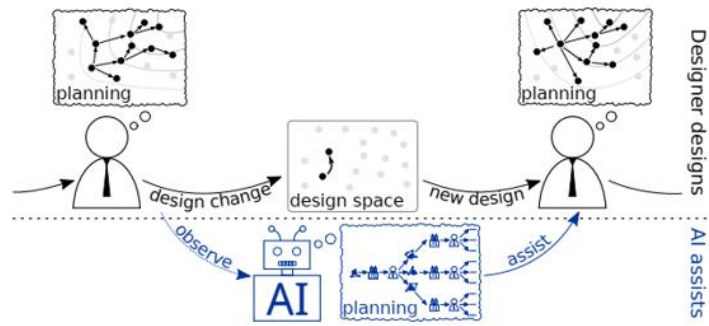

Figure 5.2  An illustration of the designer-AI collaborative design activities
on a trip planning example (De Peuter et al., 2021)

Also, Chen et al. (2019) proposed an integrated approach for enhancing design ideation by applying AI and data mining techniques. This approach consists of two models, a semantic ideation network and a visual combination model, which inspire semantically and visually based on computational creativity theory. The semantic ideation network provokes new ideas by mining knowledge across multiple domains. A generative adversarial networks model is proposed for generating UI objects for the visual combination model. An implementation of these two models was developed and tested, indicating that the approach can create a variety of cross-domain concept associations and advance the ideation process quickly and easily.

Liao et al. (2020) also proposed a framework of AI-augmented design support that involves the human ideation components and design tools related to AI in the early design stages. The framework describes the explicit roles of AI in design ideation as representation creation, empathy trigger, and engagement. The framework suggests approaches to assist cognitive patterns in the design process. An empirical study was conducted to investigate the cognitive patterns of design representations and design rationales. The study involved 30 designers with concurrent think-aloud protocols and behavior analysis. The study identified the opportunities for AI to support human creativity, and AI could provide inspiration, inform design scope, and request design actions.

### 5.3.4  Challenges and future work

Despite attempts to integrate HCI and AI, these HCI/UX designers experience challenges in incorporating AI into common UX design paradigms (Policarpo et al., 2021). We summarize the overall challenges of applying AI in HCI/UX design.

First, the AI-based approach challenges the typical activity of UX/UI prototyping. It is often difficult to convince leadership to commit to more innovative designs. The AI-based method requires an unwieldy amount of data to create a functional prototype. This approach could conflict with UX mantras like "fail fast, fail often" (Dove et al. 2017). Consequently, it isn't easy in research to experiment with many different design solutions in searching for the best. In practice, designers could not demonstrate or validate their designs' value through a working prototype as they traditionally did. Recent research also founds a lack of research integrating UX and AI (Abbas et al., 2022). One example of the obstacle is UX designers' struggle when collaborating with data scientists. Another obstacle is the lack of the tools and abilities needed to sketch or prototype when using AI as a design material.

Second, many AI-based research projects' impact remained within the academic research community and haven't succeeded in making practical influences on industry practices (Jiang et al., 2022). For instance, there is a lack of research on ML algorithms and UX, especially in envisioning how ML might improve UX (Bertão & Joo, 2021). Bridging this gap requires research that identifies practitioners' specific needs and provides translational resources to benefit from the latest technological advances and academic research findings.

Third, Abbas et al. (2022) argue that I/ML has remained underutilized to assist designers and has yet to be fully integrated into design patterns, education, and prototype tools. Therefore, tools are still in the early stages and cannot cover all conceivable questions. Also, tools were not designed with the participants of UX designers (Abbas et al. (2022).

Finally, the target users of many AI-based prototyping tools are mostly software developers rather than HCI/UX designers (Sun et al., 2020). Many HCI/UX designers lack AI knowledge, so it is still challenging to use these tools for prototyping. Further research is needed on prototyping tools for HCI/UX designers to help quickly build prototypes, discover design problems early, and reduce product development risk.

As we look forward, with the emergence of new AI-based design paradigms, HCI/UX design activities require the support of new tools. The development of these new tools should fully consider the knowledge background and way of thinking of HCI/UX designers and the collaboration between these designers and AI/software engineers in a real design environment (Sun et al., 2020). Undoubtedly, AI technology can't replace creative HCI/UX designers since these human professionals have unique capabilities to set the foundation for UX design. Still, AI definitely can support these designers in UX design as a new design material and collaborator for co-creative HCI/UX design.

## 5.4   AI for enhancing UX

### 5.4.1 Intelligent UI

AI technology is also transforming traditional UI into intelligent user interfaces (IUI).  Traditional UI techniques (e.g., mice, keyboards, and touch screens) require the user to provide inputs explicitly. AI-based approaches are now robust to inherent ambiguity and noise in real-world data to analyze and reason about natural human behavior (e.g., speech, motion, gaze patterns, or bio-physical responses). AI-based techniques can also learn high-level concepts such as user preference, user intention, and usage context to adapt the UI and proactively present information (Gebhardt et al., 2019). As a paradigm shift, AI technology holds great promise in shifting how we interact with machines from an explicit input model to a more implicit interaction paradigm in which the machine observes and interprets our actions (Li, Kumar, et al., 2020).

The idea of introducing intelligence to HCI and UI sprouted decades ago in the form of intelligent computer-assisted instructions, which later gained a wider following and application as IUIs (Maybury, 1998). IUI aims to improve the efficiency, effectiveness, and naturalness of human interaction with machines by representing, reasoning and acting on models of the user, domain, task, discourse, and media. Different disciplines support the field, including AI, software engineering, HCI, human factors engineering, psychology, etc.

Different from traditional UI, IUI should be able to adapt its behavior to other users, devices, and situations (Gonçalves et al., 2019). A non-IUI considers an "average user" in design, i.e., the UI is not designed for all types of users but for an "average" of all potential users (Ehlert, 2003). Typically, we have one context of use in non-IUI; but the context of use can change over time in an IUI. IUIs use adaptation techniques to be "intelligent/adaptive" with the ability to adapt to the user, communicate with the user and solve problems for the user" (Ehlert, 2003). Its difference from traditional UI is that they represent and reason concerning the user, task, domain, media, and situation (Jaquero, 2009).

The goal of developing IUI is to make full use of advanced AI technology to provide natural and effective human-machine dialogue. For example, new technologies (e.g., language recognition, facial recognition, gesture input, gaze tracking) offer a natural UI for systems; multi-channel interaction through multiple modalities captures user intent, behavior, and contextual scenarios to improve further the naturalness, accuracy, and effectiveness of interaction. At the same time, the effective user-centered design method is used to optimize the design of IUI.

In addition, there are several reasons why we need to research IUI. First, IUI helps promote the development of new AI technologies. Throughout modern technology development, GUI and mouse have promoted the popularization of personal computer technology; multi-touch screen technology has improved mobile phone and mobile user experience. Therefore, IUI research will find suitable application scenarios for developing AI technology.

IUI also helps further exert human capabilities and enhance human intelligence. For example, brain-computer interface research helps develop human potential and enhance the abilities of disabled people with disabilities through rehabilitation therapy. IUI actively understand user status (e.g., physiology, psychology, intention), so it will better understand users and predict their needs and behaviors, adaptively support users' activities, and ultimately make users more comfortable and interact with machines efficiently and securely.

Lastly, IUI aims to provide benefits to users such as adaptivity, context sensitivity, and task assistance (Gonçalves et al., 2019). IUI research will provide more natural and efficient intelligent systems, bringing economic benefits and considerable returns on investment to users, developers, and manufacturers. Effective IUI can improve users' work efficiency and make their work and life more convenient. The productivity of HCI can be significantly enhanced not only by contact (mouse pointing device, joystick, touchpad, keyboard, etc.) methods but also by contactless (speech and gesture commands, head and body movements, facial expressions, user's look direction, etc.) ones (Karpov & Yusupov, 2018).

Historically, interaction paradigms have guided UI development in HCI work, e.g., the WIMP (window, icon, menu, pointing) paradigm. However, WIMP's narrow sensing channels and unbalanced input/output bandwidth restrict human-machine interaction (Fan et al., 2018). Table 5.3 summarizes the new HCI characteristics that AI technology has brought in, emerging human factors issues, and critical issues for future HCI/UX work (Xu, 2019, 2020).

Table 5.3  New characteristics and human factors issues of AI technology, critical issues for HCI/UX work

| Transformative Characteristics of AI technology) | Emerging Human Factors Issues in AI Technology | Critical Issues for HCI/UX Work |
| --- | --- | --- |

| From "one-way" to "man-machine collaboration-based two-way" UI | • AI systems no longer passively accept user input and produce expected output according to fixed rules<br>• AI-based agents can actively perceive to capture and understand the user's physiological, cognitive, emotional, intentional, and other states and actively initiate interaction and push services to users | • Human-machine teaming/collaboration-based interaction models and paradigms<br>• Cognitive models of the user's states (e.g., situation awareness, physiology, cognition, emotion, and intention) |
|---|---|---|
| From "usability" to "explainable AI" UI | • AI "black box" effects can lead to inexplicable and incomprehensible system outputs<br>• AI "black box" effect raises AI trust issues | • Innovative UI technologies (such as visualization) and design<br>• "Human-centered" explainable and understandable AI (Ehsan et al., 2021)<br>• Accelerated transformation of psychological explanation theories |
| From "simple attributes" to "contextualized" UI | • AI system input includes "contextualized" data (e.g., the context of usage, user behavior), besides traditional information (e.g., simple objects such as target location and colors) | • Modeling and intelligent deduction of "situational" features (e.g., user characteristics, digital personas) based on data such as operating context and user behavior<br>• Personalized functionality suitable for user needs and usage scenarios |
| From "precise input" to "fuzzy reasoning," UI | • User input is not just a single precise form (e.g., keyboard, mouse), but may also be multimodal, ambiguous interactions (e.g., user intent)<br>• Ambiguous interaction issues in operating scenarios (e.g., random interaction signals and ambient noise) | • Methods and models for inferring user interaction intentions under uncertainty<br>• Naturalness and effectiveness of HCI in an ambiguous state |
| From "interactive" to "collaborative" UI | • UI supporting both human-machine interaction and human-machine teamwork<br>• UI supporting effective human-machine collaboration | • Alternative design paradigms and models for human-machine collaborative UI<br>• UI design standards for intelligent HCI<br>• Interaction design effectively supports human-machine collaboration (e.g., human-machine control hand-over in an emergency) |

As Table 5.3 lists, these transformative characteristics of AI technology lead to the need for innovative UI capabilities and interaction paradigms (e.g., two-way, collaborative UI). This will ultimately prompt the development of more natural and effective IUI and will require HCI/UX professionals to develop more effective approaches to explore the design of innovative UI design.

From a methodology perspective, research shows that there is a lack of effective methods for designing IUI, and HCI/UX professionals have had difficulty performing the typical HCI activities of conceptualization, rapid prototyping, and testing (Yang et al., 2020; Holmquist, 2017; Dove et al., 2017). The HCI/UX community has realized the need to enhance existing methods (Stephanidis, Salvendy, et al., 2019; Xu, 2018; Xu & Ge, 2020). To this end, Xu, Dainoff, et al. (2021) assessed existing methods of HCI, human factors, and other related fields. As a result, they proposed alternative approaches that can support the effective design of IUI better. These alternative methods can help HCI/UX professionals overcome the limitations of conventional HCI methods when designing IUI.

From a process perspective, research shows HCI/UX professionals have challenges integrating HCI/UX processes into the process of developing IUI systems. For instance, many HCI/UX professionals joined AI projects only *after* the requirements were defined (Yang, Steinfeld et al., 2020). Consequently, the design recommendations from HCI/UX professionals could be quickly declined (Yang, 2018). AI professionals often claim that many problems that HCI could not solve in the past have been solved through IUI technology (e.g., voice UI), and they can design the interaction by themselves. Still, studies have shown that the outcomes may not be acceptable from a UX perspective (e.g., Budiu & Laubheimer, 2018). Some HCI/UX professionals find collaborating effectively with AI professionals challenging due to a lack of a shared process and a common language (Girardin & Lathia, 2017). Also, studies have shown that HCI/UX professionals are not prepared to provide effective design support for AI systems (Yang, 2018).

For future work, we offer several strategic recommendations. Firstly, HCI/UX professionals need to integrate HCI/UX methods into the development process of IUI to maximize interdisciplinary collaboration. For instance, to understand the similarities and differences in practices between HCI/UX professionals and other professionals, Girardin & Lathia (2017) summarize a series of touch points and principles. Within the HCI community, researchers have indicated how the HCI/UX process should be integrated into the process of developing IUI systems (Lau et al., 2018). Specifically, Cerejo (2021) proposed a "pair design" process that puts two people (one HCI/UX professional and one AI professional) working together as a pair across the development stages of IUI systems.

Secondly, HCI/UX professionals must update their skillsets and knowledge in AI. While AI professionals should understand HCI/UX approaches, HCI/UX professionals also need to have a basic understanding of AI technology and apply the knowledge to facilitate the process integration and collaboration so that HCI professionals can fully understand the design implications posed by the unique characteristics of AI technology and be able to overcome weaknesses in the ability to influence IUI systems as reported (Yang, 2018).

Thirdly, future work needs to adapt AI technology to human capability. Human-limited cognitive resources become a bottleneck of HCI design in the pervasive computing environment. For instance, in an implicit interaction scenario initiated by intelligent ambient systems, intelligent systems may cause competition between human cognitive resources in different modalities, and users will face a high cognitive workload. Thus, HCI design must consider the "bandwidth" of human cognitive processing and resource allocation while developing innovative approaches to reduce user cognitive workload through appropriate interaction technology, adapting AI technology to human capabilities.

Fourthly, we need to develop new interaction paradigms that better fit IUI. IUI requires effective UI paradigms. In the realization of IUI, hardware technology is no longer an obstacle, but the user's interaction ability has not improved. Designing effective multimodal integration of sight, hearing, touch, gestures, and other parallel interaction paradigms is an essential part of HCI research in the age of intelligence. Historically, interface paradigms and models have guided the development of human-computer interaction (e.g., WIMP). However, the limited perception channels and unbalanced input/output bandwidth of WIMP restrict the further evolution of the UI in the AI age. Existing studies have proposed the concepts of Post-WIMP and Non-WIMP, but the effectiveness remains to be further verified. HCI/UX community should support defining paradigms, metaphors, and empirical validation to solve unique problems in IUI. It requires HCI/UX professionals to explore innovative ideas that can effectively facilitate interaction in IUI.

Finally, we need to develop HCI design standards that specifically support the development of IUI. Existing HCI design standards are primarily grown for non-IUI, and there is a lack of design standards and guidelines explicitly supporting IUI design. IUI design standards need to consider the unique characteristics of AI technology fully. There are initial design guidelines available, such as the "Google AI + People Guidebook" (Google PAIR, 2019) and Microsoft's 18 Design Guidelines (Amershi et al., 2019). The HCI/UX community must play a key role in developing these design standards.

### 5.4.2 AI assistants

An intelligent assistant (IA) is an AI/ML-based computer system capable of intelligently assisting people. IAs have gained in popularity over recent years; it ranges from helping people develop skills and exercise properly to rehabilitate physically [Islas-Cota et al., 2022], among other application domains. IAs are being deployed across domains, such as health, education, online social services, driving, domestic environment, enterprise/industry, fitness, and learning. To perform their users' daily tasks or services, IAs can send a message, make a phone call, search for specific information, set a reminder or calendar, and provide personalized recommendations. IAs are intelligent agents that employ AI techniques to provide a human-like interface (e.g., voice, vision) (Hu et al., 2019). They are also expected to perform more complex tasks, such as making purchases and accessing or managing smart IoT devices (Han & Yang, 2018). Natural language processing and AI technologies enable IAs to self-learn users' schedule and taste through daily interactions and collecting awareness data (e.g., location and context) from the Internet of Things (IoT), and then autonomously perform tasks based on user preferences and habits (Santos et al., 2016).

The objectives of IAs are to increase efficiency in an activity, better cope with an illness, resolve a problem, support everyday situations, refine skills, and attain a healthy life. Ultimately, IAs aim to enhance UX in their daily work and life. One good example of IA is voice assistants, which have been rising recently, such as Amazon Alexa, Google Assistant, and Siri from Apple (Zwakman et al., 2021). Voice assistants help facilitate human–computer dialogue naturally and intuitively, like conversations between humans.

Islas-Cota et al. (2022) presented a systematic review aiming to classify recent advances in IAs in terms of IAs' objectives, application domains, and workings. They identified what AI/ML techniques are used to enable the AI assistants. As a result, the study proposes a taxonomy of IAs, as illustrated in Figure 5.3.

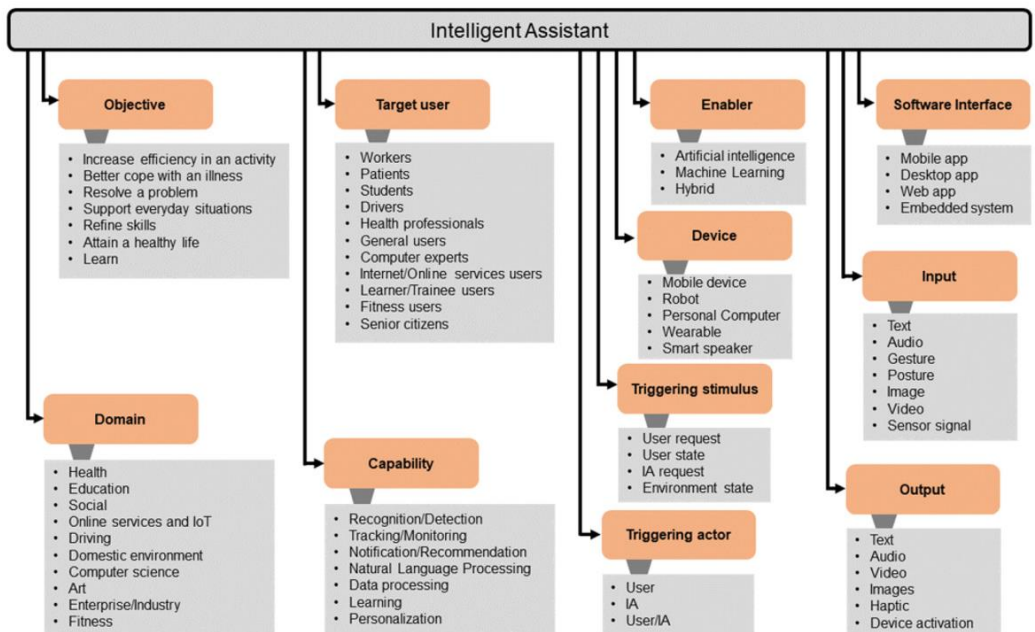

Figure 5.3  Taxonomy of IAs (Islas-Cota et al., 2022)

Research into AI-based digital assistants has a long history, dating back to Joseph Weizenbaum's well-known ELIZA in 1966 (Maedche et al., 2019). In parallel, global technology companies such as Microsoft, IBM, Google, and Amazon have been working with AI-based digital assistants to provide significant opportunities. The rise of IAs has opened a broad research area for HCI/UX professionals. It is a technology with an explicit interface to users and could therefore provide a fruitful avenue for HCI/UX research (Enhancing UX/AI assistant 5). Much research has been done, but the work primarily focused on improving the technology. Their indirect objective is to enhance the usability of the IAs, and not on the usability aspect per se. Budiu & Laubheimer's (2018) usability study found that both voice-only and screen-based intelligent assistants worked well for only minimal, simple queries with relatively simple, short answers. Users had difficulty with anything else. In addition, as part of experience issues, the privacy and security aspect of the IAs (e.g., voice assistants) still exists, as many IAs are prone to various attacks that might steal user information (Zwakm et al., 2021). There are several ways in which IAs could be used that can create new ethical and legal issues (Almeida et al., 2020).

Besides the usability design and interaction issues of IAs, the collaborative relationship between humans and AI-based IA systems is another important topic for HCI/UX professionals. Traditionally, AI-based applications are considered a tool in support of humans. We should stop thinking of AI as a developing phenomenon independent of humans, and it is necessary to move on to the consideration of hybrid intelligence. Hybrid intelligence can be further understood from three aspects, considering hybrid intelligence as the sum of human and machine efforts in achieving a goal; the amplifier of human intelligence at the physiological level; and a partnership between humans and machines (Shichkina et al., 2017).

Shichkina et al. (2017) further argued that IAs should not be just the creation of hybrid intelligence but a co-evolutionary hybrid intelligence (CHI). CHI is a symbiosis of artificial and natural intelligence, mutually developing, teaching, and complementing each other in co-evolution. Human-machine intelligence co-evolution is the fundamental building of more robust intelligent systems (Krinkin et al., 2021). Based on the concept of CHI, the goal of IAs is the mutual development of human and artificial intelligence as a single indivisible organism.

From the perspective of human-machine intelligence complementarity, the most significant potential for IAs is a mutually beneficial collaboration (Maedche et al., 2019). Both humans and machines have relative strengths. While machines are ideal for conducting repeatable, highly structured tasks, collecting, storing, processing vast amounts of data, and predicting the future in stable environments, humans can handle abstract problems and deal with fragmented information much more efficiently.

Functional and task allocation between humans and machines are a classical activity for HCI/UX professionals. There is an intensive discourse on how humans interact with AI-based technologies and how the performance of a particular task should be divided between these two entities. It is crucial to involve humans to an appropriate level in task performance, depending on the task characteristics and the context (Maedche et al., 2019). A significant challenge for future research is to investigate how to distribute the tasks between these two entities at an appropriate level to achieve optimal performance.

Autonomy is another topic that little research has been done to examine the issue from the perspective of autonomy as intelligent agents (Hu et al., 2019). In the past, many topics about autonomy focused on human autonomy, such as job autonomy, human autonomy, and community autonomy. IA autonomy refers to the fact that IA, as an intelligent agent, can independently complete tasks in some scenarios. Autonomy is a double-edged sword factor for IAs, as it can increase benefits (e.g., exerting specific risky tasks) (Robert & You, 2018). It may allow a machine to over-control without human authority, which may put humans in a risk situation in some domains (Xu, 2019, 2020).

The over-emphasis on human autonomy is because machines were not smart enough in the past. They can be regarded as relatively automated rather than autonomous. Recently, AI-enabled IAs to self-learn users' preferences through daily interactions and personal data, which ensures intelligent agents' autonomy (Hu et al., 2019). For instance, IAs will set the alarm clock to wake up according to the user's habits, reflecting the IA scheduling autonomy. Still, the complexity of IA executing both instructions is hidden, which will result in the user losing control over the specific execution process. Hu et al. (2019) raised several research questions for future work of AIs, such as whether decision-making autonomy will have a positive influence on perceived competence, whether perceived competence will have a positive effect on the intention of a user to IA continuous usage, and whether perceived uncertainty will harm the purpose to IA continuous usage.

We summarize the future research directions for improving the UX of IAs (Islas-Cota et al., 2022; Maedche et al., 2019; Zwakman et al., 2021).

- *Understanding human users' needs of IA.* Ensure that the design of IAs is in line with the user population's and society's goals and values. Research is needed to create a rich understanding of the needs and usage of the potential users of such assistants

- *Interaction technology*: Need to improve the technology empowering these IAs to provide better capabilities, such as voice recognition, the ability to understand multiple languages, providing human-like speech output, adding emotions to these devices, and likewise
- *User privacy*: Ensure that the users can trust them in their daily usage because many IAs collect potentially sensitive data from users' activities, such as visited locations
- *Collaboration:* Need to explore further how to design effective collaboration between humans and IAs. Technically speaking, the agent-based paradigm of IAs supports collaborative problem-solving either with other agents or with agent-human teams. Future work needs to exploit further the agent paradigm where multiple agents (namely, IAs) can coordinate, collaborate, and negotiate among themselves to provide users with a multi-domain ubiquitous assistance
- *Evaluation:* Need to evaluate the overall system performance from a collaboration perspective. A common practice is comparing with utilized benchmark datasets to evaluate their IAs. We need to find practical evaluation approaches to assess the efficiency and effectiveness of IAs in a collaborative way
- *Functional and task allocations*: Need to investigate from a conceptual perspective the interplays between humans and machines when using IA, investigating design variants of IA for different task types. Collaboration between humans and IA may depend on the task type to achieve optimal experience.
- *Context-aware assistance*: Exploit unsupervised ML techniques to discover users' context and behavioral patterns. IAs can provide users with personalized and contextual assistance by establishing a context. Features such as users' activities, interactions, status, and intent detection, can establish a context that enables IAs to determine how and when assistance should be provided.
- *Emotionally aware assistance:* Need to explore more elaborate emotional models. Emotions are critical to humans in decision-making and communication, among other everyday activities.
- *Virtual, augmented, and mixed reality*: Need to leverage these technologies to help improve user experience. Currently, there is a lack of IAs taking advantage of virtual, augmented, and mixed reality. For instance, IA can use virtual reality devices to assist patients with physical rehabilitation and train surgeons. Further HCI work is needed to enhance user experience while interacting with IA.
- *Human characteristics*: Need to assess human characteristics in the design of IA, such as user expertise with the technology, personality, culture, social norm, delivering personalized assistance

### 5.4.3 Recommender systems

Recommender Systems (RS) are software tools that support human decision-making, especially when choices are made over large product or service catalogs (Ricci et al., 2021). Recommender systems are integral to many of today's websites and online services. After 30 years, personalized recommendations are ubiquitous, fueled by advances in AI technology. To a large extent, making recommendations is an HCI/UX topic, which aims to determine how a computerized system can effectively support users in information search or decision-making contexts for an optimal experience.

The academic and industrial communities have proposed many recommender software and algorithms (Elahi et al., 2021). Most of these algorithms can gather various data types and exploit them to generate recommendations. These data types can describe either the item content (e.g., category, brand, and tags) or the user preferences (e.g., ratings, likes, and clicks). A recommendation list for a specific user is then made by filtering the items representing similar features to the rest of the items that the user liked/rated high. However, users may be exposed to risks, such as bad user experience and decision difficulty. If the set of recommendations is unfortunate (e.g., poor decisions, the pre-selection of items, or the decision bias ), this might lead to a poor experience.

The goal of a recommender system is to predict user interests and infer their mental processes. For example, personalized recommender systems are one of the most widely used fields of big data technology. It is implemented by mining user attributes through user behavior data and realized through the inference of basic information (e.g., user's age, gender, residence, and educational background based on the user's online browsing behavior) (Wang et al., 2013). Based on the results of the inferred data (e.g., digital personas, user profile, behavior, preference), systems can intelligently send recommendations to target users with a personalized experience. As a success case, Amazon's recommendation engine provides it with a conversion rate of up to 60% and a sales contribution rate of 30% (Li et al., 2015).

In general, there are two traditional recommender systems (Elahi  et al., 2021)

- *Collaborative filtering*: The method predicts users' preferences (i.e., ratings) by learning the preferences that a group of users provided and suggests to users the items with the highest predicted priorities. It is used in almost all application domains and relies on big data of ratings acquired from a typically extensive network of users (Desrosiers & Karypis, 2011). The underlying assumption is that users with similar preferences will also have similar preferences in the future.

- *Content-based***:** Content-based methods adopt content-based filtering (CBF) algorithms to build user profiles by associating user preferences with the item content (Deldjoo & Atani, 2016). Content-based approaches recommend items that share characteristics with items that the user has previously liked (e.g., items with a similar description or genre) (Calero Valdez et al., 2016; Zangerle et al., 2022).

  Typical fields of application are recommending movies, music, or related products in e-commerce.

Despite these traditional methods' effectiveness, as we enter the age of big data and AI, more advanced techniques have been developed to build intelligent systems for quicker, more accurate, and personalized recommendations tailored to each user's needs and preferences (Elahi et al., 2021).

There are several types of AI-enabled recommender systems that have been explored in academia and the industry:

- *Data-driven recommendations:* This method enables leveraging ML technologies to contextualize the big data to enhance the precision of suggestions, which facilitates the use of content (Beheshti et al., 2020). The approach moves from traditional statistical modeling to advanced AI-based models, which will improve mining patterns between items and user descriptors to build better suggestions.

- *Knowledge-driven recommendations:* This method empowers simulating the expertise of the domain experts (e.g., crowdsourcing methods) and adopting techniques such as reinforcement ML to enhance the system's capability for making relevant and accurate recommendations (Beheshti et al., 2018).
- *Conversational recommendations:* This method provides more sophisticated interaction paradigms for preference elicitation, item presentation, or user feedback through conversational interactions between users and recommender systems (Lei et al., 2020).
- *Intelligent ranking-based recommendations:* It can be trained by the domain experts' knowledge and experience to understand the context, extract related features, and determine the causal connections among various features over time. The goal is to change from statistical modeling to novel forms of modeling, such as deep learning, to improve potential similarities among descriptors and build a more accurate ranking (Chen et al., 2020).
- *Intelligent personalization-based recommendations:* It can support analytics around users' cognitive activities to provide intelligent and time-aware recommendations. The method tailors product and content recommendations to users' profiles and habits by analyzing users' behavior, preferences, and history. This process requires automatic data processing to identify meaningful features, select suitable algorithms, and use them for training a proper personalization model (Herath & Jayarathne, 2018).
- *Cognition-aware recommendations:* It aims to recognize the users' personalities and emotions and analyze their characteristics and affinities over time. The system needs to interpret social information (at a group level or on a one-to-one basis) and provide context-aware recommendations (Beheshti, Yakhchi et al., 2020).

Many AI models have been adapted for use in these AI-enabled recommender systems. For instance, deep neural networks for collaborative filtering to model the user-item interactions, including deep factorization machines or (variational) autoencoders (Zangerle & Bauer, 2022). Convolutional Neural Networks (CNN) are primarily used for learning features from (multimedia) sources for learning the data from audio signals (Van den Oord et al., 2013) or modeling latent features from user reviews and items (Zheng et al., 2017). Recurrent Neural Networks (RNN) are used to model sequences for sequential recommendations (Quadrant et al., 2017). Reinforcement learning models incorporate user contexts while continuously updating and optimizing the recommendation model based on user feedback (Zheng et al., 2018).

Early research on recommender systems focuses on algorithms and their evaluation to improve recommendation accuracy (Calero Valdez et al., 2016). After a few decades, the field of recommender systems has been driving toward consensus; that is, accuracy only partially constitutes the UX of a recommender system. As a result, there is an evolution from research on algorithms to research on UX with recommender systems (Konstan & Terveen, 2021).

Human-centered recommender systems are an approach that focuses on understanding the characteristics of recommender systems and users as well as the relationships between them. The goal is to design recommender systems' algorithms and interactions to fulfill better users' goals (Konstan & Terveen, 2021). Different from traditional technology-centered recommender systems, a different set of questions need to be answered from the human-centered recommender systems. For instance, what does it mean for a recommendation to be good? How many products are too many to recommend? Should I show the best recommendations or save some for later? When should the recommendations be diverse concerning each other or the user's history? What type of recommendations leads to better UX? Konstan & Terveen (2021) presented HCI research work focusing on UX and interactive visualization techniques to support the transparency of results. In addition, there is also a need for frameworks to combine human-centered recommender systems research with the best ML algorithms to achieve scalable, efficient human-centered recommender systems (Konstan & Terveen, 2021).

Furthermore, to enhance user experience, Ekstrand et al. (2014) conducted research to understand how users perceive their performance, including dimensions of accuracy, diversity, novelty, personalization, and satisfaction. The work shows that these factors should be included in an analysis of algorithm performance while building a structural equation model. Willemsen et al. (2016) studied user choice overload in the context of recommender systems. The results show that the diversity of items recommended affected the effort required to make a choice; diversity led to higher satisfaction choices but not always the highest-scoring choices for users. Research also shows that a recommender system built to optimize user engagement (rather than predictive accuracy) leads to recommendations that increase subsequent user engagement compared to predictive accuracy recommenders (Zhao et al., 2018).

How to effectively measure the UX of recommender systems is essential, so HCI/UX professionals can identify the pain point to close the gap in design. The performance of recommender systems is typically evaluated using offline and online experiments (Zangerle & Bauer, 2022). When assessing the effectiveness of a recommender system, people largely adopt offline rather than live user studies methods. Conversely, real users are requested to evaluate the recommendations in online studies. Offline studies are more popular than user studies, which are more complex and time-consuming. However, measuring the UX of recommender systems is often challenging. Konstan & Riedl (2012) argue that evaluating the UX requires a broader set of measures. Regular algorithmic work can be done by using existing datasets; measuring UX requires developing additional capabilities that include both algorithms and UI.

More specifically, Knijnenbur et al. (2012) propose a user-centric approach for evaluating recommender systems. The framework links objective system aspects to accurate user behavior through a series of perceptual and evaluative constructs. It also incorporates the influence of personal and situational characteristics on the UX. The framework was validated using a method called structural equation modeling. The results show that subjective system aspects and experience variables are invaluable in explaining why and how the UX of recommender systems comes about; the perceptions of recommendation quality and variety are essential mediators in predicting the effects of objective system aspects on the three components of UX: process (e.g., perceived effort, difficulty), system (e.g., perceived system effectiveness) and outcome (e.g., choice satisfaction). Also, the study finds that these subjective aspects strongly correlate to user behaviors (e.g., reduced browsing for higher system effectiveness).

Based on current literature, the following list summarizes the suggestions for future HCI/UX work in designing recommender systems (Calero Valdez et al., 2016; Konstan & Riedl, 2012; Konstan & Terveen, 2021; Jannach et al., 2021):

- *Better understand how users make decisions:* For example, we need to know how recommender systems can adapt to different needs (e.g., new users vs. experienced users) and how they can balance short-term with longer-term value.
- *Putting the user in control*: Users are often more satisfied when given control over how the recommender system functions. We need to design for the sweet spot so that the recommender system can balance serving users effectively while the users have the desired control.
- *Developing adaptive recommender systems:* Previous research shows that user satisfaction does not always correlate with high recommendation accuracy and the user's knowledge level and interests. There is a need to adapt recommender systems and their user interfaces to these other personal and situational characteristics.
- *Supporting affective design:* Emotions play a crucial role in human decision-making. Future work needs to explore novel sensing technologies for capturing user behavioral data (e.g., physiological data, facial expressions, speech) so that recommender systems can detect emotions and adapt recommendations based on emotional responses.
- *Conducting ongoing research:* For instance, the research on real applications that allow incorporates diverse contexts, including multi-interaction modalities (e.g., voice/audio vs. text vs. visual interaction) and decision nature (e.g., health/habit, low- vs. high-stakes)
- *Developing rigorous methods for evaluating UX:* We need to continue to adopt rigorous methods for assessing the UX and user satisfaction, such as the structural equation modeling
- *Designing for high-risk domains:* Spending money on an undesired product is the most significant risk for a user of e-commerce sites. Risk-aware algorithms or predictions of risks need to be further investigated. For instance, how to effectively visualize or communicate the uncertainty and risk of a recommendation to users, which is crucial for the systems in high-risk domains (e.g., medicine).
- *Developing insightful "beyond-accuracy" measures*: Many current methods rely on data-centric "offline" experiments that do not involve the human in the loop. We should focus much more on how systems affect both organizations and entire experience journeys, the diversity of the recommendations, or the novelty of the identified items.
- *Developing integrated solutions for better UX*: A fundamental challenge to the field of recommender systems is the integration of content-based approaches (e.g., product information, user profiles), collaborative approaches (e.g., explicit and implicit ratings, tagging, and user preference), and contextual approaches (e.g., business rules, location, user task and mood, UI ) into comprehensive recommender systems.

## 5.5 Conclusions

AI technology is transforming how HCI and UX professionals work towards delivering optimal UX in their solutions, including all aspects of user research, UI technologies and design, and UX evaluation. AI-based solutions have raised user and academic awareness of technical innovation. As a result, AI is becoming increasingly popular in improving the quality of UX. This chapter summarizes how AI technology can help HCI/UX professionals in HCI/UX research and evaluation, HCI/UX design, and enhancing UX by leveraging AI-based capabilities. It also highlights the benefits of deploying AI technology for HCI/UX activities, the challenges that HCI/UX professionals face, and future HCI/UX work.

To push the boundaries of what AI might be and might do, we need to continue to identify major unknown topics as a basis for future research endeavors (Yang, 2018). We must bridge the gap between HCI/UX professionals' work practices and AI-enabled capabilities. Professionals across disciplines need to seize this opportunity to create something entirely new. It's essential to treat the adoption of AI as a significant strategic and cultural shift for improving UX, not simply the installment of new technology (Schwartz et al., 2018). We need to collaborate on developing new methods, tools, and processes to help HCI/UX and AI professionals better innovate with AI (Bertão & Joo, 2021).